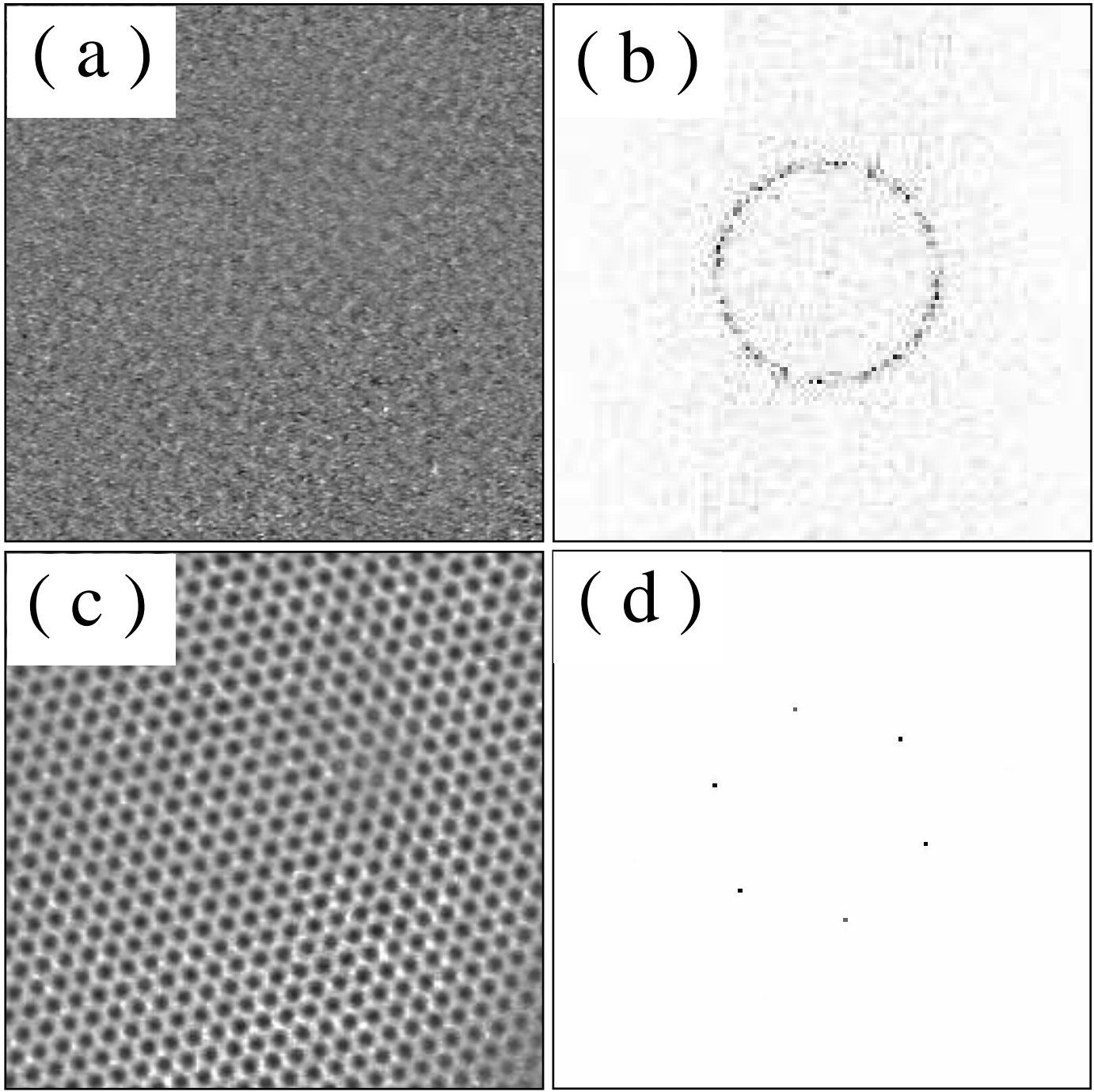

Fig. 1  Wu et. al. 94

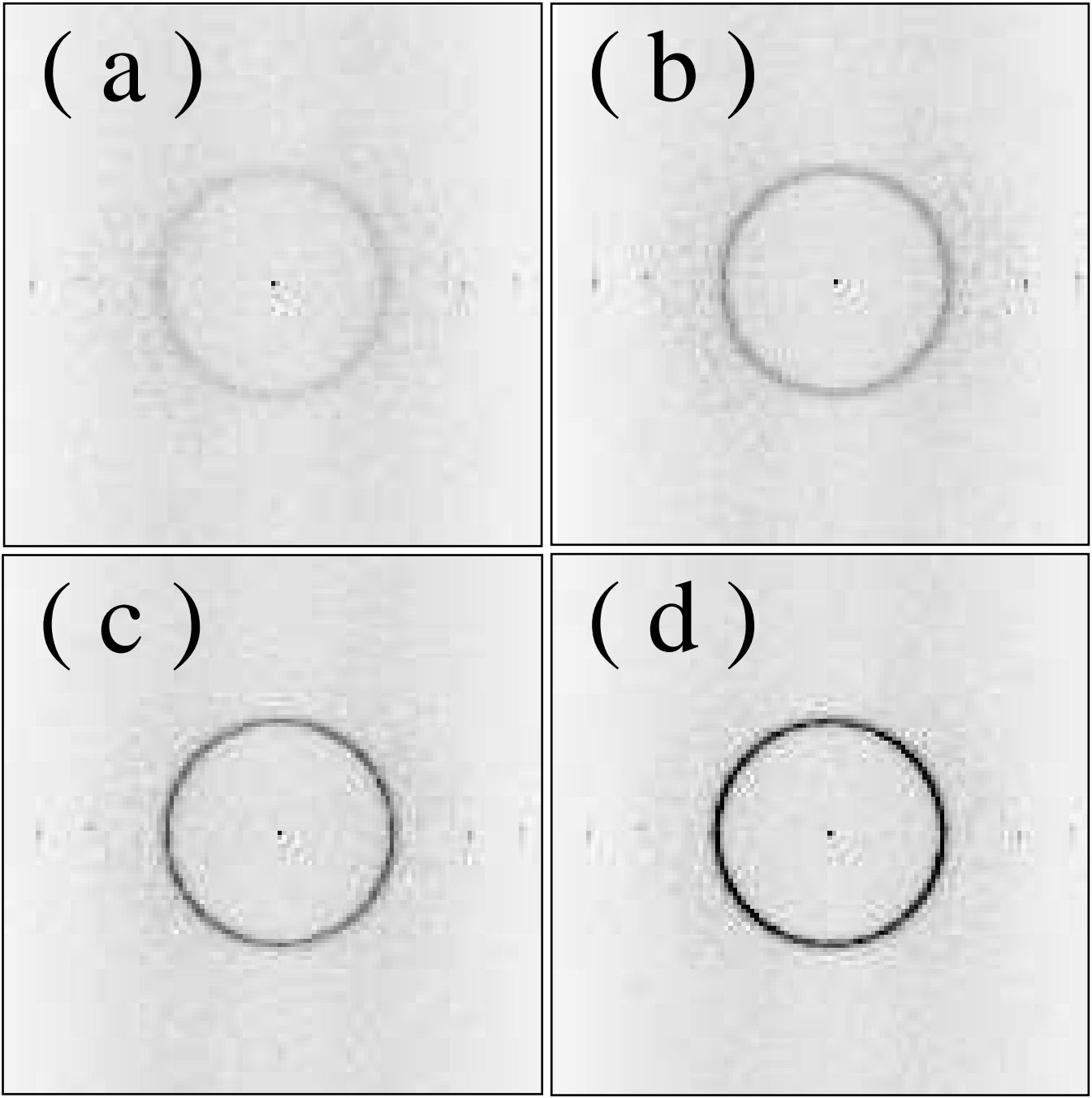

Fig. 2, Wu et. al., 94

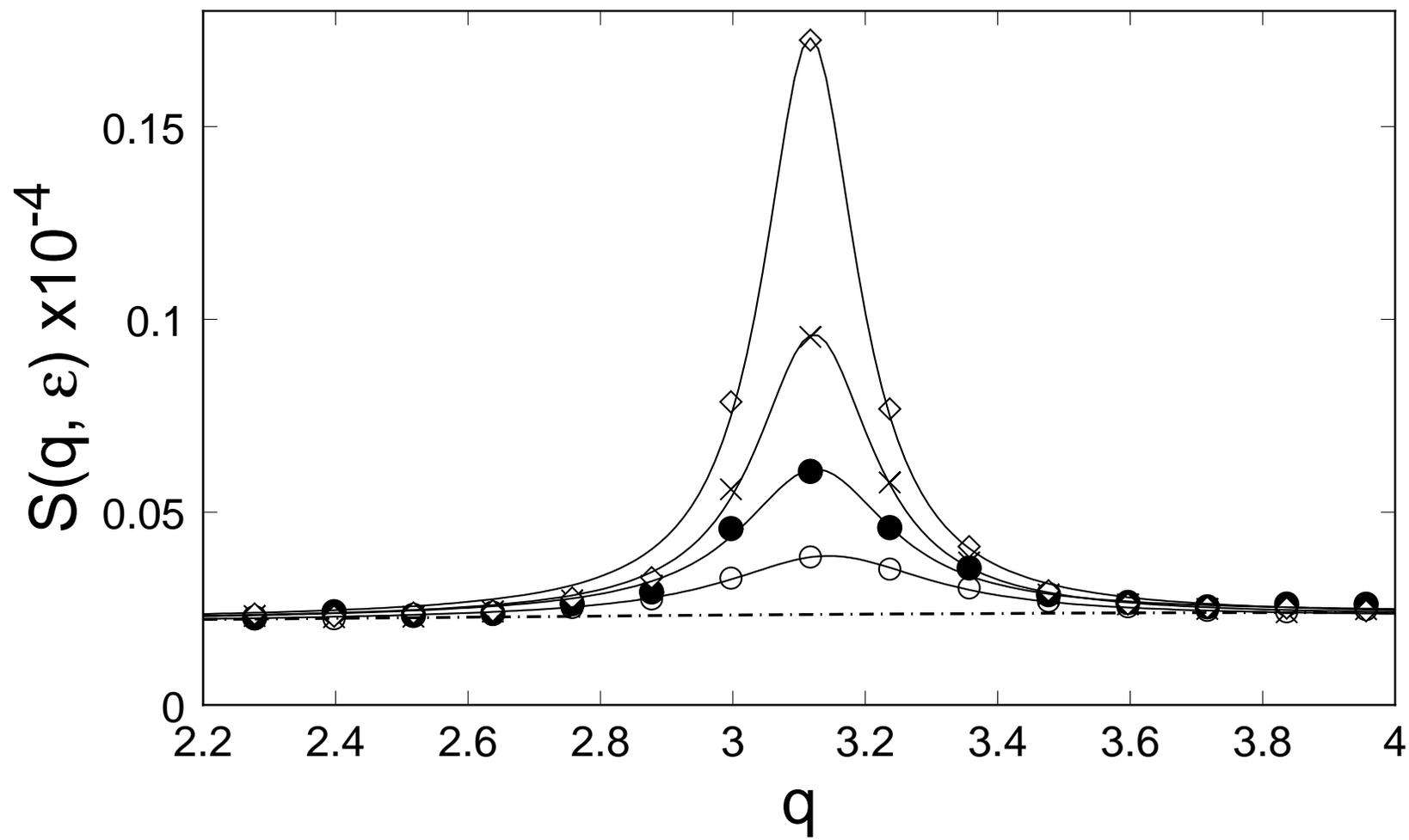

Fig. 3, Wu, et al. 94

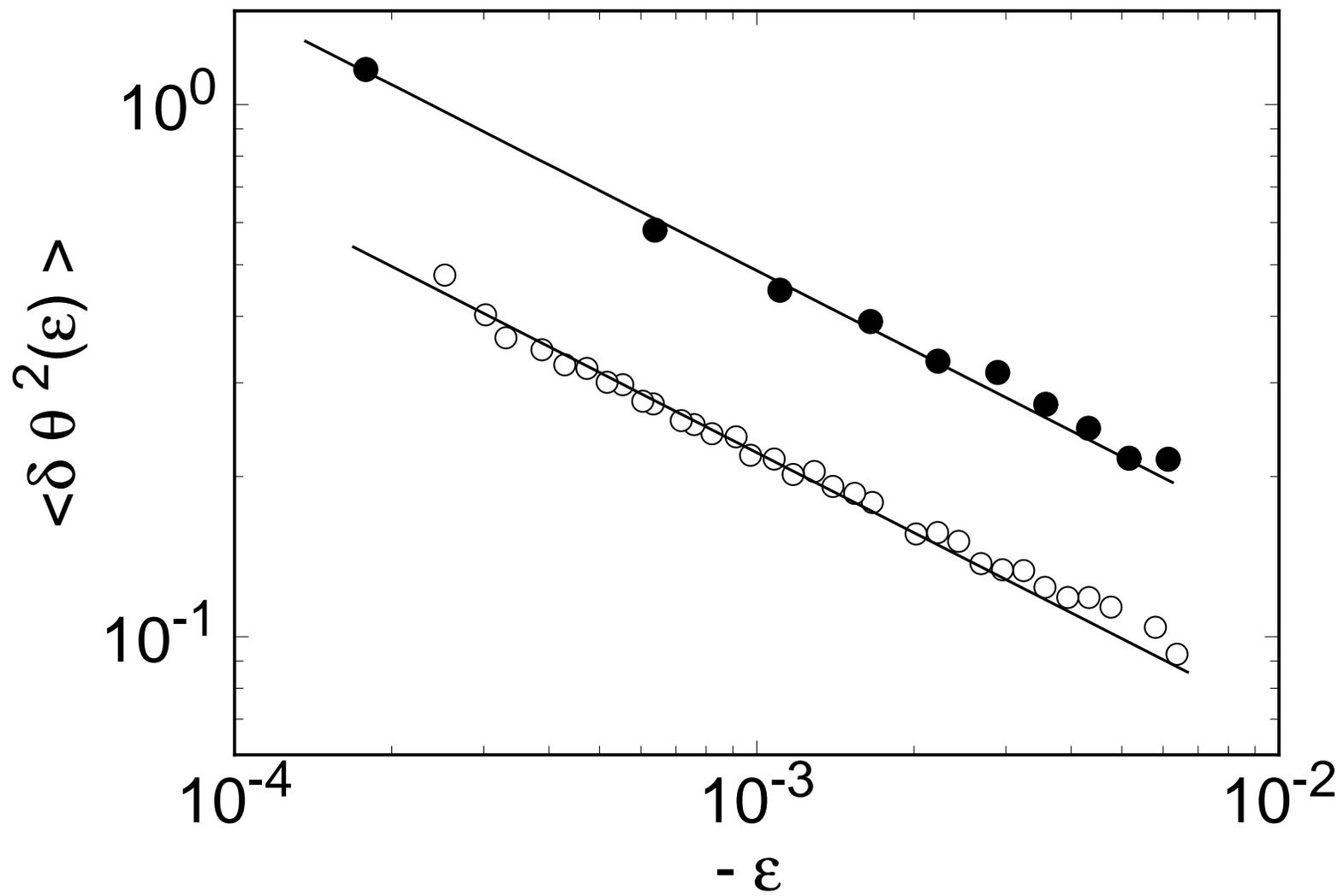

Fig. 4, Wu, et al. 94

FIGURE CAPTIONS

Fig. 1. Grey-scale images. (a): Shadowgraph image of fluctuating rolls, at a pressure of 28.96 bar, for $\epsilon = -3.0 \times 10^{-4}$. (b): Square of the modulus of the Fourier transform of the image in (a). (c): Shadowgraph image of a hexagonal pattern, at a pressure of 28.96 bar, for $\epsilon \simeq 0$. (d): Square of the modulus of the Fourier transform of the image in (c).

Fig. 2. Grey-scale images of the structure factor for gas convection in $CO_2$ at a pressure of 28.96 bar at each of four $\epsilon$ values. (a): $\epsilon = -4.2 \times 10^{-3}$. (b): $\epsilon = -1.6 \times 10^{-3}$. (c): $\epsilon = -7.1 \times 10^{-4}$. (d): $\epsilon = -3.0 \times 10^{-4}$.

Fig. 3. Azimuthal integral $S(q, \epsilon)$ of the structure factor $S(\mathbf{q}, \epsilon)$ as a function of $q$ for various $\epsilon$ at a pressure of 28.96 bar. $\diamond$: $\epsilon = -4.2 \times 10^{-3}$. $\times$: $\epsilon = -1.6 \times 10^{-3}$. $\bullet$: $\epsilon = -7.1 \times 10^{-4}$. $\circ$: $\epsilon = -3.0 \times 10^{-4}$.

Fig. 4. The variance $\delta\theta^2(\epsilon)$ of the temperature fluctuations as a function of $\epsilon$ on logarithmic scales. $\bullet$: P = 42.33 bar. $\circ$: P = 28.96 bar. The solid lines are fits of $\delta\theta^2(\epsilon) = A/\sqrt{-\epsilon}$ to the data.

21. P.C. Hohenberg and J.B. Swift, *Phys. Rev. A* **46**, 4773 (1992).

22. J.B. Swift and P.C. Hohenberg, private communication.
12

| P (bar) | $\Delta T_c^{expt}$ | $10^7 F^{expt}$ | $10^7 F^{th}$ | $\mathcal{P}^2$ |
|---|---|---|---|---|
| 28.96 | 23.56 | 1.79 | 2.31 | 6.3 |
| 28.96 | 23.56 | 1.89 | 2.31 | 6.3 |
| 31.02 | 17.27 | 2.42 | 2.48 | 3.9 |
| 42.33 | 5.46 | 4.15 | 4.09 | 0.96 |

The authors wish to thank Eberhard Bodenchartz, Pierre Hohenberg, Stephen Morris, Steve Trainoff, and Maurice M.H.P. van Putten for valuable discussions. This work was supported by the Department of Energy through Grant No. DE-FG03-87ER13738.

of the cell. They also computed (Eq. 16 of BC) the quantity $\mathcal{N} - 1$, where the Nusselt number $\mathcal{N}$ is the ratio of the effective conductivity in the presence of hydrodynamic flow to the conductivity in the absence of flow. Replacing the sum obtained by BC by an integral and making the approximation of Eq. (5), HS evaluated the result of BC for the limit of an infinite system very near threshold. Using the scaling of the present paper, they obtained

$$\mathcal{N} - 1 = \frac{F^{th}}{4\sqrt{-\epsilon}} \quad , \tag{7a}$$

with (Eq. A22 of HS)

$$F^{th} = \frac{k_B T}{\rho d \nu^2} \times \frac{2\sigma q_o}{\xi_o \tau_o R_c} \quad , \tag{7b}$$

where $\xi_o = 0.385$. They also give the relation

$$\delta\theta^2 = \tilde{c}^2(\mathcal{N} - 1) \quad , \tag{7c}$$

with $\tilde{c} = 3q_o\sqrt{R_c} = 385.28$. In analogy to Eqs. 7 we define an experimental noise power by

$$F^{expt} = 4A/\tilde{c}^2 \quad , \tag{8}$$

with A given by the fit of Eq. 6 to the experimental data for $\delta\theta^2(\epsilon)$. In Table 1 we compare $F^{th}$ with $F^{expt}$. We note that there is a trend of $\mathcal{R} \equiv F^{expt}/F^{th}$ with the extent of the departure from the Boussinesq approximation as measured by the parameter $\mathcal{P}^2$. Within experimental uncertainty, $\mathcal{R}(\mathcal{P}^2)$ extrapolates to unity as $\mathcal{P}^2$ vanishes. Thus we conclude that in the Boussinesq limit the experimental noise power is, within our resolution, equal to the theoretical estimate[2,7] for thermal noise based on the stochastic hydrodynamic equations[1].

can be obtained by integrating the structure factor of the shadowgraph signal

$$\delta\theta^2(\epsilon) = \mathcal{A}^{-2}\int[S(\mathbf{q},\epsilon) - S_B(\mathbf{q})]d\mathbf{q} = 2\pi\mathcal{A}^{-2}\int_0^\infty q[S(q,\epsilon) - S_B(q)]dq \quad . \quad (4)$$

For the structure factor given by Eq. (2), the integral in Eq. 4 diverges at large $q$. The problem is attributable to the truncation involved[22] in deriving Eq. 2. In their evaluation of the results of BC, HS made the approximation

$$\int_0^\infty \frac{q\,dq}{(q-q_o)^2 - \epsilon/\xi_o^2} \simeq q_o \int_0^\infty \frac{dq}{(q-q_o)^2 - \epsilon/\xi_o^2} \quad (5)$$

(see Eq. A16 of HS) in order to avoid this problem. Equation (5) can be justified[22] to lowest order in $\epsilon$ on the basis of the exact result of Zaitsev and Shliomis.[2] It is clearly a good approximation at small $\epsilon$ where $S(q,\epsilon)$ has a small width, and yields

$$\delta\theta^2(\epsilon) = A/\sqrt{-\epsilon} \quad . \quad (6)$$

We shall make the same approximation in the determination of $\delta\theta^2$ from the experiment, and expect that corrections of higher order in $\epsilon$ will cancel to a large extent in the comparison with theory. Figure 4 gives the results as a function of $\epsilon$ at two pressures. The solid lines are fits of Eq. (6) to the data. The adjustable parameters were the amplitude $A$ and $\Delta T_c$. The results for $\Delta T_c$ were typically very slightly larger (by a few parts in $10^4$) than the value at which a hexagonal pattern first appeared, suggesting a slightly premature transition in the presence of the fluctuations. The statistical errors derived from the fits were typically a few percent, but we expect that systematic errors from various sources may increase the uncertainty of $A$ to about 10 %.

The amplitudes of the fluctuating modes below but close to onset were calculated by BC[7] (Eqs. 9, 10b, and 12b of BC), using no-slip boundary conditions at the top and bottom though deterministic contributions have been eliminated, it still has a smooth $\epsilon$-independent background $S_B(q)$ due to camera noise which exists even for $\Delta T = 0$. We obtained $S_B(q)$ by fitting the experimental data of $S(q)$ vs. q well away from $q_o$ to a quadratic in $q^2$. The solid lines are fits to the function

$$S(q,\epsilon) = \frac{I_0}{(q - q_o)^2 + \Gamma^2} + S_B(q) \quad (2)$$

which is expected to pertain[7,21] close to threshold, and the background $S_B$ is shown as the dashed line. Equation (2), with $I_0$, $q_o$, and $\Gamma$ adjustable, was found to provide an excellent fit to the data for all pressures and $\epsilon$ studied.

For comparison with theory, the quantity of interest is the mean square amplitude of $\theta(\mathbf{x}, z, \epsilon)$, the fluctuation in the temperature field. As was done by Hohenberg and Swift[21] (HS), we scale temperature by $(\kappa\nu/\alpha g d^3) = \Delta T_c/R_c$ ($R_c = 1708$ is the critical Rayleigh number) and length by $d$, and write $\theta(\mathbf{x}, z, \epsilon)$ as $\theta(\mathbf{x}, \epsilon)\tilde{\theta}_0(z)$. As in HS, the vertical eigenfunction $\tilde{\theta}_0(z)$ is normalized so that its square integrates to unity (see Eq. A24b of HS).

For our experimental setup the shadowgraph signal $\delta I(\mathbf{x}, \epsilon)$ and the temperature fluctuation $\theta(\mathbf{x}, \epsilon)$ are directly proportional[18,19]

$$\delta I(\mathbf{x}, \epsilon) = \mathcal{A}\theta(\mathbf{x}, \epsilon). \quad (3)$$

The constant $\mathcal{A}$ can be written in dimensionless form as $2\gamma(z_1)q_o^2 z_1 (\partial n/\partial T) <\tilde{\theta}_0(z)>_z$. Here $z_1$ is the optical distance from the cell to the imaging plane, $n$ is the refractive index of the fluid, the vertical average $<\tilde{\theta}_0(z)>_z$ is equal to 0.8892, and $\gamma(z_1)$ is a numerical factor which may be computed[19] on the basis of physical optics, and which for our geometry is equal to 0.81. Consequently the mean square amplitude of the fluctuations $\delta\theta^2(\epsilon) = <\theta^2(\mathbf{x}, \epsilon)>_\mathbf{x}$ deterministic flow velocity grows as $|\epsilon|^{-1/2}$ whereas that of the fluctuations grows only as $|\epsilon|^{-1/4}$. Consequently, near the transition even a microscopic dust particle can force flow in the form of concentric rings[10] which may contribute to a shadowgraph image. In the parameter range of interest the velocities are so small that the system is linear. Thus superposition is valid, and the deterministic signal could be identified by averaging all the signal images $I_i(\mathbf{x}, \epsilon)$, taken at a given $\epsilon$. It could then be removed by subtracting this average from each signal image taken at that $\epsilon$. Figure 1a shows a grey-scale rendition of such a difference image $\delta I(\mathbf{x}, \epsilon)$, for $\epsilon = -3.0 \times 10^{-4}$. It reveals some spatial variation in excess of instrumental noise, but the detailed structure of the fluctuating field is hard to discern. Figure 1b is a grey-scale rendition of $|\delta I(\mathbf{q}, \epsilon)|^2$, where $\delta I(\mathbf{q}, \epsilon)$ is the spatial Fourier transform of $\delta I(\mathbf{x}, \epsilon)$. A dark ring is apparent, indicating that the fluctuations can be considered as superimposed convection rolls with many different orientations and a preferred wavenumber, $q_o$. Figure (1c) is taken barely above the onset (nominally $\epsilon = 0$). The image shows a defect-free hexagon pattern.[10] The modulus squared of its Fourier transform is displayed in Fig. (1d). Notice that the hexagon wavenumber is essentially the same as the radius $q_o$ of the ring in Fig. (1b).

In order to measure the mean square amplitude of the fluctuations as accurately as possible, we averaged 64 Fourier images of the sort shown in Fig. (1b), at each $\epsilon$, to give the structure factor $S(\mathbf{q}, \epsilon) \equiv < |\delta I(\mathbf{q}, \epsilon)|^2 >$. The results at several $\epsilon$ are shown in Fig. 2. The structure factor shows no obvious azimuthal variation, and thus reflects the underlying rotational invariance of the RBC system.[20] As $\epsilon$ approached zero, the rings became darker, showing that the fluctuations become stronger as the system approaches the deterministic onset. The azimuthal average of $S(\mathbf{q}, \epsilon)$, which we denote $S(q, \epsilon)$, is shown in Fig. 3. Al-the conduction state to convection in the form of hexagons occurred.[10] At 28.96, 31.02 and 42.33 bar, we found $\Delta T_c$ to be 23.56, 17.27 and 5.46 $\pm 0.002$ °C, and we changed $\epsilon$ in steps of $8 \times 10^{-5}$, $1 \times 10^{-4}$ and $4 \times 10^{-4}$, respectively.

The fluctuating flows were visualized by the shadowgraph technique.[17–19] The light beam passed through the cell twice vertically, being reflected from the bottom plate. Images contained 256 × 256 pixels and covered an area 2.5 cm by 2.5 cm. A time series of 128 background images was taken at $\Delta T_c - 2.0$ °C at time intervals large compared to $t_l$. Each of these images was the average of 16 images taken 0.44 s apart. Here the fluctuations were extremely weak, but to further reduce their contribution, the average $\tilde{I}_o(\mathbf{x})$ of the 128 images was used as the background image ($\mathbf{x}$ is the horizontal position). After this, $\Delta T$ was ramped up slowly to $\epsilon \simeq -0.006$, where the fluctuations became large enough to measure. A series of 64 images $\tilde{I}_i(\mathbf{x}, \epsilon)$ was taken (again each image was an average of 16 taken 0.44 s apart) at each of many $\epsilon$-values for $-0.006 \leq \epsilon < 0$. These were used to compute the signal image

$$I_i(\mathbf{x}, \epsilon) = [\tilde{I}_i(\mathbf{x}, \epsilon) - \tilde{I}_o(\mathbf{x})]/\tilde{I}_o(\mathbf{x}) \ . \tag{1}$$

Before each image sequence at a new $\epsilon$-value, the system was equilibrated for one hour. The time between successive images was kept approximately equal to $t_l$ so the measurements were nearly uncorrelated. In obtaining the amplitude of the fluctuations a small ($\leq 20\%$) $\epsilon$-dependent correction was made to account for the effect of averaging 16 images for each final image.

Ideally, for $\epsilon < 0$, the flow would consist only of fluctuations. However, imperfections in the cell caused *deterministic* flow. Although the imperfections were *extremely* small, their relative influence increased as the bifurcation was approached from below because the with the value calculated by van Beijeren and Cohen[7] (BC) for thermal noise with rigid boundaries.

The only previous measurement of thermal fluctuations in a hydrodynamic system suitable for comparison[8] with theory of which we are aware is due to Rehberg *et al.*,[9] and involved electro-convection in a nematic liquid crystal (NLC). Even though that system is "macroscopic", it is particularly susceptible to noise because the physical dimensions are only of order 10 $\mu$m and because the elastic constants, which determine the macroscopic energy to which $k_BT$ has to be compared, are exceptionally small. In a NLC there is a preferred direction, and thus electroconvection and RBC belong to different symmetry classes. The measurements reported here reveal the effect of the rotational invariance in the horizontal plane of the RBC system on its fluctuations.

We used a circular cell[10] filled with $CO_2$ at pressures of 28.96, 31.02 and 42.33 bar, held constant to $\pm 10^{-4}$ bar. The cell height was $d = 468$ (481) $\mu$m, and the aspect ratio (radius/height) was 85 (83) at 28.96 (31.02 and 42.33) bar. The top and bottom plates were 0.953 cm thick optically flat sapphire discs whose temperatures were held constant to $\pm 0.3$ mK. Using interferometry, we found $d$ to be constant to $\pm 0.15 \mu$m over the central 80% of the cell radius. The sidewall was made of paper. The top and bottom plate temperatures were adjusted so as to keep their mean fixed at 32.00 °C. The density[11] $\rho$, isobaric thermal expansion coefficient[11] $\alpha$, heat capacity[12] $C_P$, shear viscosity[13] $\eta$, and thermal conductivity[14] $\lambda$ are given in a footnote.[15] The vertical thermal diffusion time $t_v \equiv d^2/\kappa$ ($\kappa = \lambda/\rho C_P$) was near 1 s, and typical fluctuation life-times are given by $t_l = t_v \tau_0/|\epsilon|$ with[16] $\tau_0 \simeq 0.07$. The Prandtl number $\sigma \equiv \nu/\kappa$ ($\nu$ is the kinematic viscosity) was 0.91, 0.92 and 1.04 at 28.96, 31.02 and 42.33 bar, respectively. When $\Delta T$ exceeded $\Delta T_c$, a transcritical bifurcation from Bifurcations in spatially extended dissipative systems are usually discussed in terms of deterministic equations for the macroscopic variables which neglect thermal noise. Many such "ideal" systems undergo a sharp bifurcation at a critical value of a control parameter, where a spatially uniform state loses stability and a state with spatial variation appears. However, if noise is present, it will drive fluctuations away from the uniform state, even below the bifurcation. As near a thermodynamic critical point, the fluctuation amplitudes grow as the bifurcation is approached because the susceptibility diverges there. Using the stochastic hydrodynamic equations introduced by Landau and Lifshitz,[1] this problem was considered theoretically over two decades ago[2–4] for the case of Rayleigh-Bénard convection (RBC), which is the buoyancy-induced motion in a shallow horizontal layer of fluid heated from below. For RBC the deterministic model predicts pure conduction until the temperature difference $\Delta T$ exceeds a critical value $\Delta T_c$. In the presence of noise, time-dependent fluctuating flows are predicted to occur even for $\Delta T < \Delta T_c$. They have zero mean, but their root-mean-square amplitude is finite. This amplitude diverges at $\Delta T_c$ when nonlinear saturation is neglected. These fluctuations induced by thermal noise were expected to be unobservably weak because the thermal energy $k_B T$ which drives them is many orders of magnitude smaller than the typical kinetic energy of a macroscopic convecting fluid element.

In this Letter, we report experimental measurements of fluctuations below $\Delta T_c$ in a large-aspect-ratio convection-cell.[5,6] Using the shadowgraph technique, we observed fluctuating convection rolls of random orientation. Their structure factor consisted of a ring without significant angular variation. The mean square fluctuation amplitude was found to increase as $\epsilon \equiv \Delta T/\Delta T_c - 1$ approached zero, within experimental resolution proportional to $1/\sqrt{-\epsilon}$. These experimental results agree with predictions based on the Navier-Stokes equations with additive noise terms.[1] The noise power necessary to explain the amplitude agrees well

# Thermally Induced Fluctuations Below
# the Onset of Rayleigh-Bénard Convection
## February 9, 1995

Mingming Wu, Guenter Ahlers, and David S. Cannell

*Department of Physics and*
*Center for Nonlinear Science*
*University of California*
*Santa Barbara, CA 93106*

We report quantitative experimental results for the intensity of noise-induced fluctuations below the critical temperature difference $\Delta T_c$ for Rayleigh-Bénard convection. The structure factor of the fluctuating convection rolls is consistent with the expected rotational invariance of the system. In agreement with predictions based on stochastic hydrodynamic equations, the fluctuation intensity is found to be proportional to $1/\sqrt{-\epsilon}$ where $\epsilon \equiv \Delta T/\Delta T_c - 1$. The noise power necessary to explain the measurements agrees with the prediction for thermal noise.

PACS numbers: 47.20.-k, 43.50.+y, 47.25.Mr